\documentclass[submission,conference,creativecommons]{eptcs}
\providecommand{\event}{MARS2020} 
\usepackage{underscore}           
\usepackage{cite}
\usepackage{url}
\usepackage{amsmath,amssymb,amsfonts}
\usepackage{algorithmic}
\usepackage{graphicx}
\usepackage{subcaption}
\usepackage{comment}
\usepackage{caption}
\usepackage{algorithm}
\usepackage{textcomp}
\usepackage{xcolor}
\usepackage[export]{adjustbox}
\usepackage{listings}

\definecolor{eclipseBlue}{RGB}{42,0.0,255}
\definecolor{eclipseGreen}{RGB}{63,127,95}

\lstdefinelanguage{Gamma}
{
	morekeywords={
		true,
		false,
		const,
		boolean,
		integer, 
		natural,
		else,
		enum,
		not,
		and,
		or,
		xor,
		interface,
		extends,
		in,
		out,
		inout,
		event,
		import,
		package,
		@schedule,
		bottom-up,
		top-down,
		statechart,
		rate,
		var,
		transition,
		from,
		to,
		region,
		initial,
		shallow,
		deep,
		history,
		state,
		entry,
		exit,
		choice,
		raise,
		raised,
		set,
		assign,		
		sync,
		cascade,
		adapter,
		async, 
		of,
		port,
		provides,
		requires,
		component,
		execute,
		bind,
		->,
		channel,
		-o)-,
		any,
		run,
		full,
		step,
		reset,
		priority,
		capacity,
		clock,
		s,
		ms,
		when,
		queue,
		act,
		assert,
		states,
		elapse,
		schedule
	},
	sensitive=true, 
	morecomment=[l]{//}, 
	morecomment=[s]{/*}{*/}, 
	morestring=[b]" 
}

\newcommand{\setGammaSyntax}{
	\lstset {
		language={Gamma},
		basicstyle=\scriptsize\sffamily, 
		captionpos=b, 
		extendedchars=true, 
		tabsize=2, 
		columns=fixed, 
		keepspaces=true, 
		showstringspaces=false, 
		breaklines=true, 
		frame=trbl, 
		frameround=tttt, 
		framesep=4pt, 
		commentstyle=\color{eclipseGreen}, 
		keywordstyle=\color{black}\bfseries, 
		stringstyle=\color{eclipseBlue}, 
	}
}

\title{Simulation-based Safety Assessment of\\High-level Reliability Models}

\author{Simon J\'ozsef Nagy
\institute{Budapest University of Technology and Economics, Budapest, Hungary\\
Department of Measurement and Information Systems}
\email{simon.jozsef.nagy@gmail.com}
\and
Bence Graics  \qquad\qquad Krist\'of Marussy \qquad\qquad Andr\'as V\"or\"os 
\institute{Budapest University of Technology and Economics, Budapest, Hungary\\
Department of Measurement and Information Systems\\
MTA-BME Lend\"ulet Cyber-physical Systems Research Group}
\email{graics@mit.bme.hu \qquad\qquad  marussy@mit.bme.hu \qquad\qquad  vori@mit.bme.hu}
}
\def\titlerunning{Simulation-based Safety Assessment of High-level Reliability Models}
\def\authorrunning{Simon J\'ozsef Nagy \& Bence Graics \& Krist\'of Marussy \& Andr\'as V\"or\"os}

\begin{document}

\newcommand{\todo}[1]{\color{red}TODO: #1\color{black}}
\maketitle

\begin{abstract}
Systems engineering approaches use high-level models to capture the architecture and behavior of the system. However, when safety engineers conduct safety and reliability analysis, they have to create formal models, such as fault-trees, according to the behavior described by the high-level engineering models and environmental/fault assumptions. Instead of creating low-level analysis models, our approach builds on engineering models in safety analysis by exploiting the simulation capabilities of recent probabilistic programming and simulation advancements. Thus, it could be applied in accordance with standards and best practices for the analysis of a critical automotive system as part of an industrial collaboration, while leveraging high-level block diagrams and statechart models created by engineers. We demonstrate the applicability of our approach in a case study adapted from the automotive system from the collaboration.
\end{abstract}

\section{Introduction}

Safety-critical cyber-physical systems, such as embedded control systems in the automotive domain, must satisfy numerous stringent extra-functional requirements, such as safety, reliability, and availability, in addition to functional requirements. The ISO 26262 \cite{iso26262} standard for automotive systems requires addressing these issues during system design by following recommended development practices and demonstrating compliance. To comply with the standards and stakeholder requirements, the certification process of the system uses top-down deductive reliability modeling and analysis of the architecture and behaviors to verify safety properties.

Increasingly complex behaviors of embedded automotive systems, especially in the case of fault avoidance and other safety mechanisms, pose significant challenges in these analyses. Not only classical hardware redundancy mechanisms but also fault-tolerant sensor fusion algorithms and adaptive reconfiguration are applied. Thus, we need expressive modeling approaches to precisely describe the system behavior and faults, as well as tool support for calculating reliability metrics.

Our goal in this work is to use high-level statechart-based languages in the safety and reliability analysis process and support the reuse of engineering models defined in the system design phase, such as state machines and block diagrams.
Therefore, we developed an integrated automotive reliability modeling and analysis approach, which is able to capture the architecture and the behavior of the system using high-level engineering models. We provide a statechart language in the Gamma framework for the engineers to design not only the engineering models but also the error propagation and fault/environmental assumption models.
Our approach exploits the simulation capabilities of recent probabilistic programming and simulation advancements (e.g., \cite{hoffman2013stochasticsvi,chen2014stochastichmc,hoffman2014nonuts,bingham2019pyro}), and we provide the tool support to generate the input for these techniques.
Therefore, we can calculate the compliance of the safety requirements and also support the decision-making process in all phases of the system development process. 

We demonstrate our modeling and analysis approach in the context of a safety-critical automotive electrics and electronics (E/E) system. Nevertheless, our approach is general and may be used for other adaptive cyber-physical systems and in other domains too.

\subsection{Case study}\label{sec:casestudy}
We demonstrate our top-down modeling and analysis method in the context of a real-life example from the automotive industry, namely on an electronic control unit (\textit{ECU}) of an electronic power-assisted steering (\textit{EPAS}) unit \cite{burton2003innovationepas} developed by Thyssenkrupp Components.  Due to its critical role in the car, it has several safety functions, including adaptive reconfiguration and high redundancy in the hardware. In this case study, we calculate the probabilistic measures required by the ISO\,26262 standards. Note that the case study was slightly changed compared to the real system for protecting intellectual property. However, the changes do not substantially affect its structure and behavior.

\begin{figure}
\begin{subfigure}{0.30\textwidth}
\centering
\includegraphics[width=0.92\linewidth]{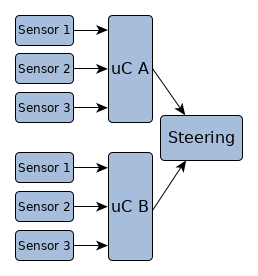}
\caption{Architecture of EPAS}
\end{subfigure}\begin{subfigure}{0.35\textwidth}
\centering
\includegraphics[width=1.00\linewidth]{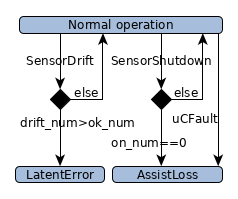}
\caption{The fault model of a uC}
\end{subfigure}\begin{subfigure}{0.35\textwidth}
\centering
\includegraphics[width=0.95\linewidth]{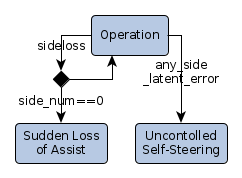}
\caption{The system states}
\end{subfigure}
\caption{Structural and behavioral models of the EPAS}
\label{fig:tkpsys}
\end{figure}

The simplified structural model of a typical, widely-used \textit{EPAS} \textit{ECU} is depicted in Figure~\ref{fig:tkpsys}a. It has two microcontrollers (uC) that are separated and can operate individually. A uC has three redundant sensors, each of which can have an operational state and two failure states, namely, \textit{Shutdown}, and \textit{Drift}. If the sensor stops due to a failure, the model enters the \textit{Shutdown} state, which can be detected every time. In contrast, the \textit{Drift} state represents a latent failure mode: in this state, the sensor seems to work correctly, yet it has an erroneous output. Thus, the detection of this failure mode requires redundant sensors and a voting mechanism. With the help of the sensors, the uCs provide the steering-actuation functionality using a closed control loop (a safety-critical function, SCF) that must operate during the lifetime of the car continuously.

Figure~\ref{fig:tkpsys}/b) depicts the simplified statechart\footnote{
Variables in the statechart:
\begin{itemize}
    \item drift num : number of drifted sensors
    \item ok num : number of normal sensors
    \item on num : number of sensors that seem to be working
\end{itemize}  } 
describing the behavior of a uC. The initial state is \textit{Normal operation}. The model goes to state \textit{AssistLoss} if the uC fails or all sensors fail (go to state \textit{Shutdown}). In contrast, the uC goes to state \textit{LatentError} if at least two sensors have latent error as the bad sensors will vote down the good one. Therefore, the uC will use wrong input data in the control loop, causing the EPAS (the states of which are modeled in an evaluation statechart) to go to state \textit{Uncontrolled self-steering}.
The EPAS system can fail in two ways as it can be seen in Figure~\ref{fig:tkpsys}/c):
\begin{itemize}
    \item If both uCs go to state \textit{AssistLoss}, the steering assistance will also stop operating, resulting in a troublesome steering experience. This state of the \textit{EPAS} is called \textit{Loss of assist} (LoA).
    \item  If a uC goes to state \textit{LatentError}, the whole EPAS will go to state \textit{Uncontrolled self-steering} (SS). This situation is extremely dangerous and must be prevented by any means, as the driver loses control over the car.
\end{itemize}

\subsection{Challenges of the safety analysis}

In the design of complex electrics and electronics (E/E) automotive systems, practitioners are well versed in classical top-down fault modeling with fault trees, as well as bottom-up analysis with failure modes, effects and diagnostic analysis (FMEDA). These techniques allow demonstrating system safety according to the applicable standards by manually creating and reviewing analysis models that are subsequently evaluated by software tools.

However, the increasing complexity of systems, such as distributed error detection and fail-over mechanisms in the EPAS, pose significant challenges for the classical approach.

Firstly, classical models may become very large and cumbersome to handle. For example, in the EPAS case study, a fault tree model required \(92\) logic gates even after simplifying the system behavior, while describing the full behavior and diagnostics is impossible with fault trees, as they do not support the definition of state-based behavior. Even though there are significantly more expressive formalisms for modeling stochastic systems~\cite{Katoen16landscape}, they often require specific expert knowledge, and cannot be applied widely by engineers. Therefore, experts must create (either automatically or by code generation) and review stochastic models based on high-level architecture models.

Secondly, different stochastic analyses are performed throughout the system design not only to demonstrate safety but also to inform design decisions and fix errors. For example, Pareto and sensitivity analysis find components most often responsible for safety goal violations, which are candidates for changes. Additionally, fitting parametric distributions in Weibull analysis~\cite{abernethy1983weibull} provides information about the product life cycle. These analyses often require purpose-built models with tailored abstractions, such as a different fault tree in the classical approach for every analysis question, which multiplies expert effort.

Lastly, the high number of hardware and software components poses a challenge for analysis tools due to the state space explosion. With more than four orders of magnitude difference between rare fault modes, iterative numerical analysis methods can be rendered ineffective, while simulation-based methods require support for rare event sampling.

These challenges made us seek an integrated modeling and analysis approach that can be applied by engineers working on the EPAS project. At the same time, it is effective enough for answering varied analysis questions throughout the development and certification process.

\subsection{Overview of the approach}

\begin{figure}
    \centering
    \includegraphics[width=1.0\linewidth]{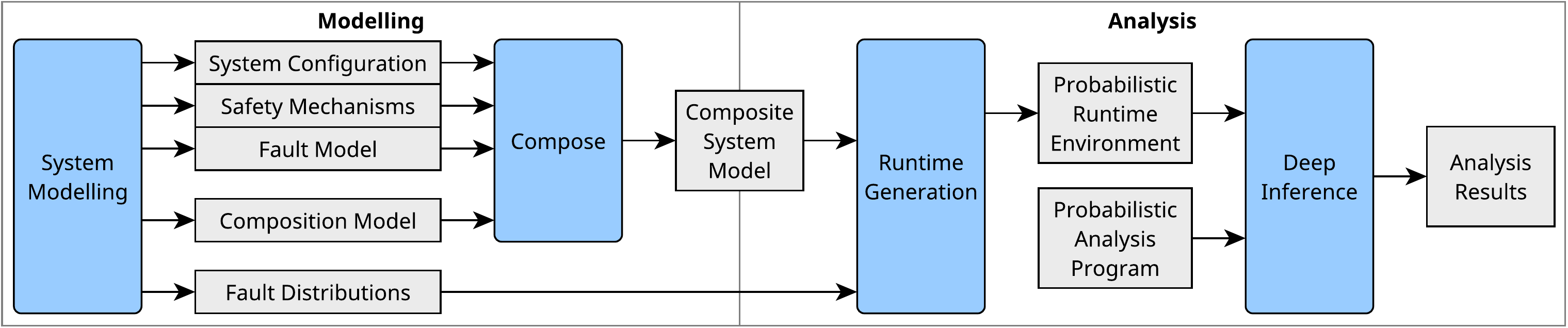}
    \caption{Activities and artifacts in the safety assessment approach}
    \label{fig:overallstructure}
\end{figure}

To overcome these challenges, we created an integrated, top-down safety modeling and analysis method (illustrated in Figure~\ref{fig:overallstructure}) for complex critical systems, such as automotive E/E components. It supports the safety assessment of the system defined by a functional model using high-level modeling languages, such as block diagrams and statecharts with which engineers are familiar.

The first step of our approach is the construction of the high-level reliability model, which specifies the failure modes of the hardware components, the behavior of safety mechanisms and the propagation of errors as interconnected statecharts.
The resulting three-layered, state-based reliability model (presented in Section~\ref{sec:statebasedfaultmodeling}) defines the possible states of the hardware components and safety mechanisms, whereas the system configuration specifies the connections between the software components as well as their interaction modes. 
Even though most of these models, such as the behavioral and structural models of the safety mechanisms are available at the beginning of the analysis in the form of design artifacts, e.g., SysML state machines and internal block diagrams, some of these models have to be created during deductive safety analysis. Thus, the same systems engineering tools can be used for creating the analysis models that are used for the design.

Secondly, the fault distributions (presented in Section~\ref{envfaultdist}) specify the stochastic behavior of the state-transitions in the state-based reliability model. Similarly to more traditional approaches, distributions are obtained from standards or FMEDA analysis.

To facilitate analysis, we developed a \textit{Probabistic Runtime Environment} (\textit{PRE}) that processes the high-level models for reliability analysis using the deep probabilistic programming~\cite{bingham2019pyro} paradigm. Various analyses can be specified as probabilistic programs, and the inference engine can handle models with large state spaces and rare events. Therefore, we can use our analysis approach both during design with complex analysis questions and for demonstrating system safety. 

To our best knowledge, this is the first application of the deep probabilistic programming to the reliability analysis of safety-critical systems.

\section{Background and related work}\label{sec:backgound}

\subsection{Safety metrics}
Critical systems have to satisfy various functional and extra-functional requirements. Dependability-related extra-functional requirements ensure that the system provides its functionalities/services steadily. In our paper, we focus on \emph{reliability} and \emph{availability} analysis. The reliability of a system represents the continuity of the correct service \cite{reliability}, and the availability of a system represents readiness for the correct service.

During safety assessment, all these concepts have to be analyzed with several techniques, e.g., lifetime prediction, which discovers when and how the system can fail. These analysis methods can produce several properties of the system, e.g., \textit{mean-time-to-first-failure} (MTFF) and the \textit{failure-in-time} (FIT). MTTF determines the average time until the first system failure, and FIT determines the probability of the system failure during a given period of time. To calculate these measures, deductive, 
top-down analysis has to be applied.

\subsection{Analysis solutions}

One of the most popular deductive analysis methods is the fault tree analysis (FTA) \cite{dugan2000developingdfta}. It supports the modeling of static systems; however, modeling components with an inner state is impossible with this method. This poses limitations for modeling reconfigurable and self-diagnosing systems.
Continuous-time Markov-chains (CTMCs) \cite{kwiatkowska2011prism} and related formalisms can model a wide range of behaviors. The explicit solution of CTMCs is hindered by the state-space explosion, i.e., the number of states becomes extremely large due to the complexity of the system-under-analysis. Various abstraction techniques were developed to tackle this challenge \cite{Katoen16landscape}, but they typically require significant modeling and analysis expertise.

Statistical model checking methods and tools, such as UPPAAL-SMC \cite{bulychev2012uppaal} were also developed: these approaches rely on random sampling to simulate the behavior of the system. The resulting data set can be analyzed using standard statistical analysis methods, e.g., statistical tests and Weibull analysis \cite{abernethy1983weibull}. UPPAAL-SMC is widely used in the railway \cite{ruijters2016betterreawayengineering}, automotive \cite{filipovikj2016simulinkautomotive} and aerospace \cite{shan2014formalaerospace} domains. Even though it can model a wide range of behaviors by providing a low-level modeling language, direct modeling of large systems may be cumbersome. Translation from higher-level models to UPPAAL is possible~\cite{molnaretalicse18}, but may substantially increase model size.

In addition, simulation-based approaches are also widely used for FTA in the field of nuclear power \cite{vcepin2002dynamicfaulttreedft}, electric power distribution systems \cite{zhang2012reliabilityftsimulation} and wastewater treatment \cite{taheriyoun2015reliabilityftsimulation}. The common disadvantage of these approaches is that they have to model every failure mode and operational mode of the system separately. As a result, they have to create a distinct fault tree for every failure mode, and they are unable to analyze the joint distribution of the failure modes. Unfortunately, the standard statistical methods are unable to analyze conditional models, even though they are used extensively during the safety assessment of critical systems.

\subsection{State-based modeling}
\label{sec:statebasedfaultmodeling}
State machines provide a convenient formalism to model the behavior of reactive systems. State machine models process incoming events and react to them in accordance with their internal states. Statecharts \cite{Harel:1987:SVF:34884.34886} are a popular extension of state machines providing complex constructions to support the high-level design of reactive systems. This formalism contains several expressive modeling elements, such as variables, arithmetic expressions, and interfaces with parametric events. 

The Gamma Statechart Composition Framework\footnote{\url{http://gamma.inf.mit.bme.hu/}} 
\cite{molnaretalicse18} is an open-source, integrated modeling toolset to support the semantically sound composition of heterogeneous statechart components~\cite{graicsmolnarminisy18}. The framework reuses statechart modeling languages of third-party tools and their code generators, e.g., MagicDraw\footnote{\url{https://www.nomagic.com/products/magicdraw}} and Yakindu Statechart Tools\footnote{\url{https://www.itemis.com/en/yakindu/state-machine/}}, for separate components. As a core element, the framework provides the Gamma Composition Language, which supports the interconnection of components hierarchically based on precise semantics. Gamma provides automated code generators as well as test case generators for the analysis of interactions between components. Gamma also supports system-level formal verification and validation (V\&V) functionalities, i.e., the system model can be exhaustively analyzed with respect to formal requirements, by mapping statechart and composition models into the input representations of verification back-ends.

\subsection{Probabilistic programming}\label{sec:probprog}

In order to formulate complex statistical inference problems, such as Bayesian machine learning and differential privacy applications as computer programs, a new programming paradigm called \textit{probabilistic programming} has been developed, which explicitly allows sampling from probability distributions as part of a program. In the last ten years, many tools have appeared to support this paradigm, e.g., Stan \cite{carpenter2017stan}, Edward \cite{tran2016edward}, Anglican \cite{andrieu2010particle}, PyMC \cite{2016ascl.soft10016Spymc3} and Pyro \cite{bingham2019pyro}.

In comparison to other tools and approaches, the greatest advantage of probabilistic programming tools is the general \emph{inference algorithm}, which can analyze conditional models and calculate posterior distributions independently from the particular formulation of the model. These conditional models are created by placing \textit{observe} statements in the program, which define the observed (conditional) distributions of random variables. By running the probabilistic program in an inference environment, we obtain the corresponding posterior distributions.

Most inference algorithms use a gradient-based Monte-Carlo method, such as the Hamiltonian Monte-Carlo \cite{chen2014stochastichmc} and the No-U-Turn algorithms \cite{hoffman2014nonuts}.
Recently, a new approach called deep probabilistic programming \cite{bingham2019pyro} has emerged relying on deep learning algorithms for fast and efficient computations. In this work, we relied on the stochastic variational inference algorithm (SVI) \cite{hoffman2013stochasticsvi} from this family, which fits a parametric distribution to the output of the probabilistic program. It optimizes the following evidence lower bound (ELBO) between the parametric \emph{guide} function $q_{\varphi}({\bf z})$ and the sampled posterior $p_{\theta}({\bf x}, {\bf z})$ (dependent on the parameters $\varphi$ and $\theta$) distribution of the output of the probabilistic program, where $\bf x$ and $\bf z$ are the conditioned and latent random variables, respectively: ${\rm ELBO} = \mathbb{E}_{q_{\varphi}({\bf z})} \left [
\log p_{\theta}({\bf x}, {\bf z}) - \log q_{\varphi}({\bf z})
\right]$.

By choosing an appropriate guide $q_{\varphi}({\bf z})$ and condition $\bf x$, different aspects of the program can be analyzed. In the probabilistic programming environment Pyro~\cite{bingham2019pyro}, both the guide, the program, and the conditions can be specified as Python programs in a user-friendly manner.

\section{Modeling the EPAS ECU}\label{overviewoftheapproach}

\subsection{State-based system modeling}

Our approach introduces a fault-modeling technique based on high-level state-based (statechart) models. We describe the system under analysis as a composition of software and hardware elements with Yakindu statecharts and Gamma modeling languages using the Gamma Composition Framework. Statecharts provide an expressive language to capture complex error propagation scenarios of systems. Therefore, in our approach, we can apply statechart-based modeling (interconnected statecharts) to describe the faulty behavior in the system, e.g., latent, cascading, multi-point, and common-cause failures.

With the expressive modeling elements of the aforementioned languages, we can model complex safety mechanisms. We use variables and arithmetic expressions to describe the complex decision-making algorithms in the reconfiguration strategies and the sensor fusion algorithms. In addition, parametric events in statechart interfaces can model both the communication and the fault propagation between the components.

We propose a top-down, three-step approach for the modeling of the EPAS system based on both the system models and safety requirements.
In Step~I, the high-level states and components of the system are defined, which is followed by Step~II, the specification of safety mechanisms. Finally, in Step~III,~elementary hardware components are modeled.

\paragraph{I. System-level behavior}

In the first step, we model the high-level state of the system from the perspective of functional and extra-functional requirements. Thus, we create the \textit{system statechart} (depicted in Figure~\ref{fig:evaluationgammastatechart1}), which contains the operating and failure states of the EPAS, namely, \textit{Normal}, \textit{Loss of Assist} and \textit{Self-steering}. 

After that, we define a \emph{composition model} of the system, which contains communication channels between the communication ports of each statechart model. Error events and reconfiguration events propagate from the lower-level components (e.g., hardware and safety mechanisms) to the higher-level components (e.g., the system-level statechart) according to the semantics of the Gamma Composition Language.

\begin{figure}[ht]
    \centering
    \includegraphics[width=1.0\linewidth]{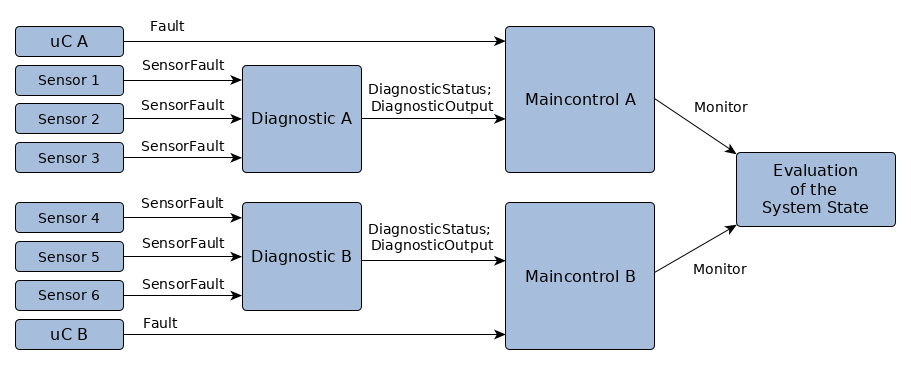}
    \caption{Layered structure of the EPAS model. Labels on arrows refer to the interfaces which define the propagated error and reconfiguration events}
    \label{fig:tkpmodelstruct1}
\end{figure}

Components are represented by statecharts, which will be acquired or defined in the subsequent modeling steps. The composed model describes the behavior of the full system, while engineers can focus on the different statechart models at the appropriate levels of abstraction for top-down (deductive) analysis. This model contains every component of the system, the failure of which can contribute to the failure of the whole system and defines how they can interact with each other. The error propagation model of the EPAS ECU is depicted in Figure~\ref{fig:tkpmodelstruct1}. 

\paragraph{II. Safety-level behavior}

In the second step, we specify the behavior of the safety mechanisms using the statechart models created by the systems engineers. In our analysis, these models were already available as part of the engineering models of the system.

\begin{figure}[ht!]
    \centering
    \includegraphics[width=0.9\linewidth]{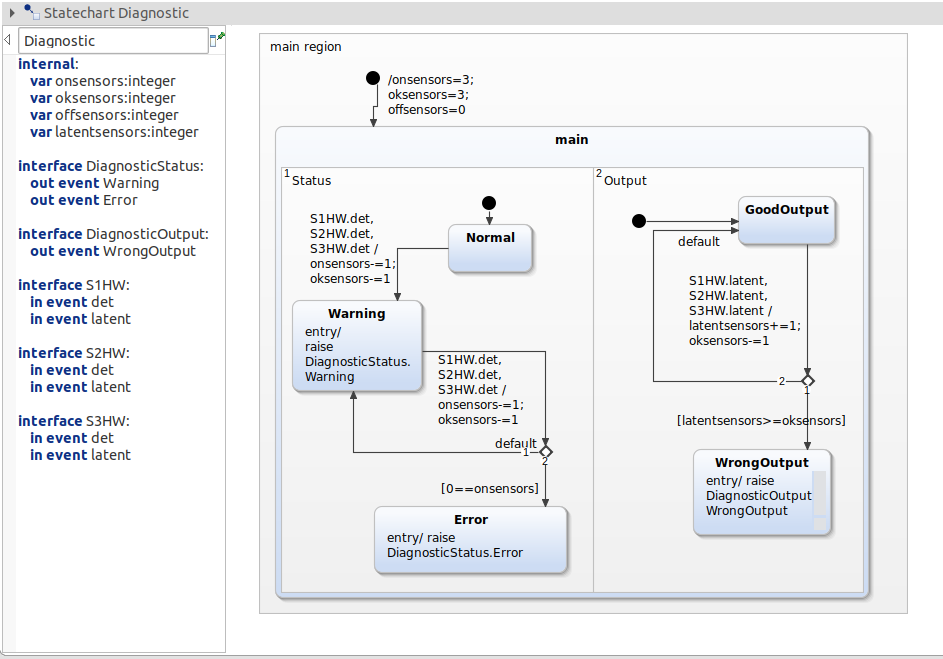}
    \caption{Statechart of the  sensor diagnostics}
    \label{fig:diagnosticgammastatechart1}
\end{figure}

The state of the EPAS ECU is determined by the state of the motor controllers. Consequently, the system statechart is connected to the statechart models of the controllers (depicted in Figure~\ref{fig:controllergammastatechart1}).
This controller switches off the actuation if either the uC is faulty or the sensor diagnostic sends an error status message. Consequently, the behavior of a controller is affected greatly by the behavior of the diagnostic function. Therefore, the controller statechart is connected to a diagnostic statechart (depicted in Figure~\ref{fig:diagnosticgammastatechart1}), which defines the voting mechanism in the diagnostic function.  Note that the expressive modeling elements of statecharts, such as orthogonal regions, variables, and arithmetic expressions greatly facilitate the specification of the complex behavior of the sensor diagnostics.

\paragraph{III. Hardware-level behavior}

In the last step, we model the elementary hardware subcomponents of the system, which are the smallest portion of hardware components considered in the safety analysis. These hardware components have an independent and distinct functionality from the perspective of the safety analysis. We can model these hardware elements with statecharts, which contain the operating and failure modes of the hardware components. Such statecharts can be created based on standards \cite{us1986military}, supplier information, and the result of other analysis methods, e.g., FMEDA.

\begin{figure}[htbp]
    \centering
    \includegraphics[width=0.8\linewidth]{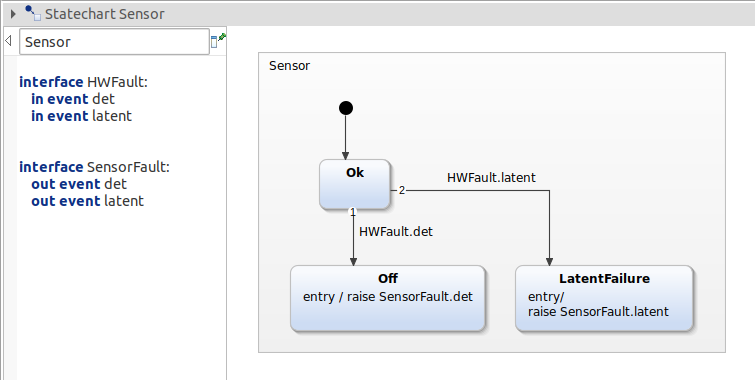}
    \caption{Statechart based failure model of the sensor}
    \label{fig:sensorgammastatechart1}
\end{figure}

\begin{figure}[htbp]
    \centering
    \includegraphics[width=0.7\linewidth]{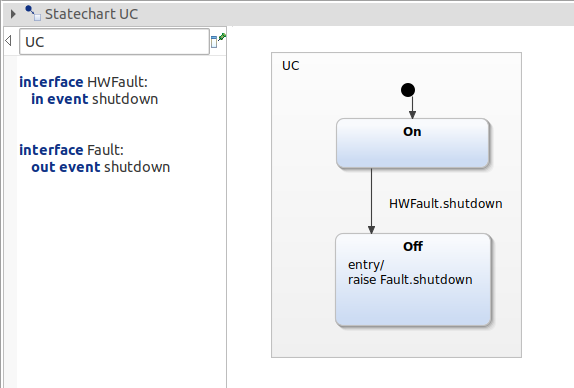}
    \caption{Statechart based failure model of the uC sensor}
    \label{fig:ucgammastatechart1}
\end{figure}

Generally, the statecharts of most hardware components, such as the uC in the ECU (Figure~\ref{fig:ucgammastatechart1}), have only two states: a \textit{Good} and a \textit{Faulty}. Although, in the case of complex adaptive systems, the failure mode greatly influences the state of the system. The sensor (Figure~\ref{fig:sensorgammastatechart1}) has an operational state and two vastly different failure states: \textit{Shutdown} and \textit{Drift}. The \textit{Shutdown} state can cause only \textit{Loss of assist} failure whereas the \textit{Drift} state is the only source of the \textit{Self-steering} failure.

\subsection{Composite model}

The statechart models introduced in Steps I, II, and III are composed into a single, deterministic, state-based model of the system according to the composition model defined in Step I.

The semantics of the composition are defined by the Gamma framework and are similar to composition facilities provided by SysML block diagrams, Extended Timed Automata, and tools like UPPAAL, albeit with support for ports, interfaces, and various synchronous and asynchronous communication semantics. For a description of the semantics, we direct the interested reader to~\cite{molnaretalicse18,graicsmolnarminisy18}, and Appendix~\ref{app:gamma}.

\begin{figure}[h]
    \centering
    \includegraphics[width=1.0\linewidth]{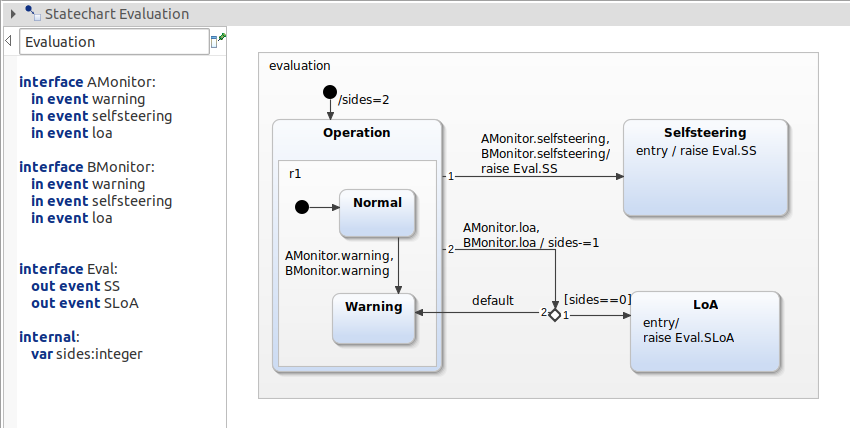}
    \caption{System level statechart of the EPAS}
    \label{fig:evaluationgammastatechart1}
\end{figure}

\begin{figure}[ht!]
    \centering
    \includegraphics[width=0.9\linewidth]{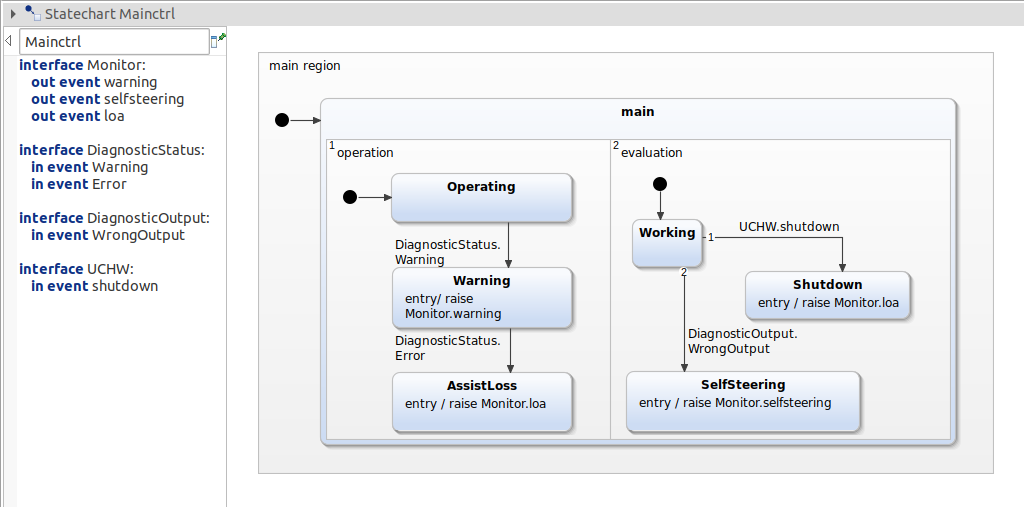}
    \caption{Statechart of the motor controller}
    \label{fig:controllergammastatechart1}
\end{figure}

As an example of event propagation in the composite model of the system, consider the transitions to the states named \emph{SelfSteering} in the system level statechart in Figure~\ref{fig:evaluationgammastatechart1} and motor controller in Figure~\ref{fig:controllergammastatechart1}. Entry to the \emph{SelfSteering} state in any of the two instances of the motor controller statechart, which correspond to the two redundant uCs of the EPAS, raises the event \emph{selfsteering} on their respective \emph{Monitor} ports. The system-level statechart listens to these events on its \emph{MonitorA} and \emph{MonitorB} ports, which are connected to outputs of the two controllers. Any of these events trigger the transition from the \emph{Operation} state to the \emph{SelfSteering} state, and raise the \emph{SS} event on the \emph{Eval} port, signalling the safety goal violation.

\subsection{Fault distributions}\label{envfaultdist}

In order to model not only the functions and services of the system but also its extra-functional aspects (including reliability and availability), we have to affix fault occurrence distributions to the low-level hardware faults in our model.

The Yakindu and Gamma modeling languages cannot express stochastic behaviors. Therefore, we model the stochastic transitions separately, by annotating them in an external table (see Table~\ref{tab:faultdistributions}). During simulation-based analysis, the distributions are sampled to generate (timed) fault sequences, which are input to the composite state-based model as a sequence of low-level hardware events. Safety goal violations can be ascertained as output events from the model.

\section{Analysis of the system}

In this section, we present an analysis method for deterministic, state-based composite models annotated by fault distribution, which can be obtained using the methodology in the previous section.

Our analysis is based on deep probabilistic programming. While probabilistic programs, such as the PRISM guarded command language, are widely used in reliability analysis, to our best knowledge, this is the first use of an inference engine with \emph{observe} statements for conditional reliability analysis and distribution fitting.


\subsection{Probabilistic Runtime Environment}\label{probruntimeenv}

In order to analyze the composite model of the system with randomly sampled fault sequences using deep probabilistic programming, we created a bridge between the state-based model and the probabilistic programming environment, called the Probabilistic Runtime Environment (PRE). This environment allows running the state-based model as part of a deep probabilistic program. The PRE can be executed on its own to compute the mean lifetime of the system, or as part of a probabilistic program containing \emph{observe} statements to facilitate conditional analysis and distribution fitting for inference.

PRE is implemented in Python to be compatible with the most popular stochastic analysis tools \cite{carpenter2017stan,bingham2019pyro,2016ascl.soft10016Spymc3}. The implementation is built on top of the Pyro language, since Pyro includes the inference algorithms needed for the analysis (summarized in Section~\ref{sec:probprog}), as well as state-of-the-art inference tools, such as GPU accelerated deep probabilistic programming.

\subsection{Translation to a probabilistic program}

Discrete models are defined in the Gamma framework, out of which a Java implementation is generated. This implementation can evaluate the system-level effects of the hardware faults via interface functions. The model of the system is then directly translated into a probabilistic program that simulates random hardware faults and the system behavior.

Our approach is based on continuous-time, asynchronous simulation. The probabilistic program generates the component failure events randomly for all failure modes of all components in the system at once. Thus, we sample each failure-mode distribution for each component and arrange the resulted events in chronological order to produce a stochastic event series for the discrete model, which determines the failure-time and failure-mode. However, the number of possible failure combinations grows exponentially with the number of hardware components in the system. Pyro provides several algorithms that can mitigate this problem, such as interest sampling and SVI.

The pseudo-code of the probabilistic simulator program is shown in Algorithm~\ref{alg:generatefaultseriespseudostatechart}. This algorithm consists of the following steps:
\begin{enumerate}
    \item First, the fault distribution of every failure mode of every hardware component is sampled and collected into the fault set.
    \item Then, the fault events are arranged in chronological order to get fault series. This procedure is essential for our reliability assessment since, during a simulation, each fault event is sent to the model of the software components individually.
    \item Finally, the result of a component failure series is evaluated in a \textit{while} loop. If the system reaches a failure state during the evaluation cycle, the \textit{while} lopp stops, and the simulation returns with both the failure state and the elapsed time to reach the failure state.
\end{enumerate}

All of these steps are integrated into the \textit{simulate} function of the \textit{PRE}. As a result, the generated probabilistic program is available for the analysis scripts, which can both validate the system requirements and identify the weak points in the system.

\subsection{Stochastic analysis of the case study}
\label{sec:analyzis}

\subsubsection{Lifetime prediction}

To predict the time to failure (TFF) of the system, the probabilistic program in Algorithm~\ref{alg:generatefaultseriespseudostatechart} can be run repeatedly to sample from the TFF distribution and estimate the MTFF by stochastic model checking methods~\cite{bulychev2012uppaal}. The results were visualized in a histogram (see Figure~\ref{fig:histograms}).

To reveal the aging of the failures of the EPAS, we performed Weibull analysis \cite{abernethy1983weibull} using stochastic variational inference (SVI) from Pyro as outlied in Section~\ref{sec:probprog}. We set $q_{\{\eta,\beta\}} (\mathit{time})=\mathit{Weibull}(\eta,\beta;\mathit{time})$ as the guide function, where $\mathit{Weibull}(\eta,\beta;\mathit{time})$ is the pdf of a Weibull distributed random variable with $\eta$ and $\beta$ scale and shape parameters respectively. In this case, the set of conditioned variables is empty, i.e., ${\bf x} = \emptyset$, since we use neither conditioning nor the \textit{observe} statement.

The SVI optimization algorithm in the \textit{PRE} explores the optimal Weibull shape ($\beta$) and scale ($\eta$) parameters, which reveals the fitted lifetime distribution. This method helps to understand the changes in the failure rate of the system over time, as indicated by the shape parameter $\beta$. 

\subsubsection{Conditional analysis}

For safety assessment, we can examine even complex, low-probability events in the system-under-analysis, such as the MTTF in case of a specified failure mode.
By utilizing the Pyro deep probabilistic programming algorithm inside the PRE, we are able to analyze joint as well as conditional distributions.
We can use all the inference algorithms implemented in Pyro (e.g., importance sampling, Hamiltonian Mont-Carlo, and SVI) to analyze efficiently complex conditional and parametric distributions even with low occurrence probability \cite{perov2019multiverse}.

To apply these inference algorithms, we first have to create a posterior model in the \textit{PRE}, namely, a \textit{conditional simulation model}. In this model, we assume that some random variables in the simulation will have a specified value (e.g., we assume that the system will go to a specified failure mode state). In order to create such a model, we put \textit{observe} statements for each conditioned random variable in the generated \textit{simulate} function. Thereafter, we either put the \textit{conditional simulation model} into a Pyro inference algorithm and run Monte Carlo simulations or we use the model fitting (SVI) algorithm of Pyro (introduced in the previous subsection) with an appropriate \textit{guide} function.

The conditional analysis can be used for component sensitivity calculations. The main objective of this method is to investigate how the lifetime of the system changes if a given component fails. We created a \textit{conditional model} where we assume (\textit{observe}) that a given component fails during the mission time. Thus, the weaknesses of the system design could be identified and remedied.

\section{Evaluation}\label{sec:evaluation}

\subsection{Compatibility with ISO 26262 safety analysis requirements}

The safety analysis of critical components has to be approved by several industry-specific standards. Therefore, we defined our analysis method in accordance with the ISO 26262-2018 standard as it is one of the most modern and relevant standards. In the following, we present the conformity between our method and the ISO 26262:
\begin{itemize}
    \item Our analysis technique supports system development in the full life-cycle, including the examination of new ideas with modular analysis even when the system model is incomplete.
    \item If the safety requirements are not fulfilled, we can construct a conditional analysis, which can identify the weaknesses of the system design. Moreover, with the help of the Gamma framework, we are able to functionally analyze the components of the system separately.
    \item The use of our analysis method does not require any special competences. Thus the safety analysis can be an integrated part of the development process. As a result, statechart-based component models can be reviewed directly by the engineers, and the engineers by themselves are able to create component models without any training.
\end{itemize}

\subsection{Analysis of the EPAS ECU}

\subsubsection{Lifetime prediction}
\label{sec:lifetime-prediction}
We ran 10,000 independent simulations in the \textit{PRE} and visualized the results in Figure~\ref{fig:lifetimedistfitcomhist} with the blue histogram. As can be seen, the results resemble the characteristics of the Weibull distribution. Thus, using the Pyro optimizers and SVI, we can fit a Weibull distribution to the system behavior, as shown in Listing~\ref{lst:weibull}. We ran 10,000 steps with the SVI algorithm to fit the model. The comparison of the real and the fitted Weibull model is depicted in Figure~\ref{fig:lifetimedistfitcomhist},
where the fitted Weibull model is represented with the orange chart. As can be seen, the resulting model is not perfect, but it gives a good approximation for the scale and the shape parameters. These parameters give an insight into how the system changes over time. In addition, the parameters may be used for system-level analysis, including all subsystems of the car.

To validate the results, we also conducted a fault tree analysis (FTA) manually on the EPAS model. As it is depicted in Figure~\ref{fig:lifetimedistfitcomhist}, the results of the FTA matched those of simulation closely.

\subsubsection{Conditional lifetime prediction}
\label{sec:conditional-lifetime-prediction}
The ISO26262 standard requires a detailed analysis of the failure modes, therefore we have to analyze the \textit{conditional behavior}, the \textit{Uncontrolled self-steering} and the \textit{Loss of assist} failure modes from the perspective of the expected occurrence time.

\begin{figure}
    
    \begin{subfigure}{0.47\textwidth}
        \centering
        \includegraphics[width=0.99\linewidth]{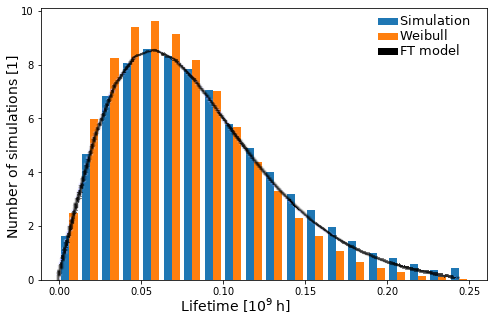}
        \caption{Comparison of the simulated and fitted lifetime distribution}
        \label{fig:lifetimedistfitcomhist}
    \end{subfigure}
    \begin{subfigure}{0.47\textwidth}
        \centering
        \includegraphics[width=0.99\linewidth]{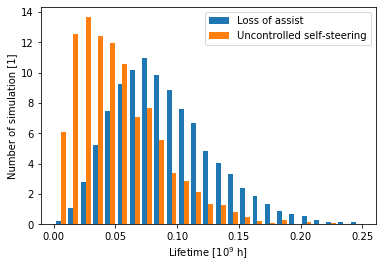}
        \caption{Comparison of the lifetime distribution of the failure modes}
        \label{fig:compcondlifetimedisthist}
    \end{subfigure}

    \caption{Lifetime prediction and conditional lifetime prediction with simulation}
    \label{fig:histograms}
\end{figure}

We created two conditional models in the \textit{PRE} that give us the posterior distribution of the lifetime, assuming that the failure mode is \textit{Loss of assist} or \textit{Uncontrolled self-steering}. The associated Pyro probabilistic program is illustrated in Listing~\ref{lst:conditional}.
We ran the inference algorithm 10,000 times. The comparison of the two resulted posterior lifetime distributions are depicted in Figure~\ref{fig:compcondlifetimedisthist}. The results meet the expectations, \textit{Uncontrolled self-steering} occurs much earlier than \textit{Loss of assist} due to the following reason: if a sensor has a latent failure and any other sensor has any kind of problem, the system will go to state \textit{Uncontrolled self-steering} immediately. The \textit{Loss of assist} failure mode occurs when both sides have a uC fault or three sensors have \textit{Shutdown} type faults. Reaching \textit{Loss of assist} takes a much longer time than \textit{Uncontrolled self-steering}.

\subsection{Run-time measurements of the algorithms}
In order to validate our analysis method, we made several run time measurements. We examined the time-to-failure analysis as well as the conditional time-to-failure analysis. We ran all analysis scripts five times and calculated the median and checked the consistency of the results. 

We evaluated our analysis method on several different models. We used the EPAS, defined in Section~\ref{sec:casestudy}, as a template of the benchmark models. Thus we created three new versions of this model. Each new model version was extended with some additional sensors and some new uCs. Similarly to the original EPAS model, each uC has a voting mechanism for their sensors, and the system goes to state \textit{Loss of assist} if no uC is operational and goes to state \textit{Uncontrolled self-steering} if at least one of them uses bad sensor-data for the steering control.
For the measurements, we used an average PC configuration\footnote{Ubuntu 18.04.3 LTS; AMD Athlon X4 860K 
CPU; 8 GB RAM; GeForce GTX 1050Ti; Pyro-0.4.1; PyTorch-0.4.1}. 
Median running times of the analyses are presented in Table~\ref{tab:run-time-measurements}.

The results show that the run-time of a single simulation scales well as the size of the model increases and the analysis method can be applied even for large models. 

\begin{table}[t]
    \centering
    \begin{tabular}{r||c|c|c|c}
        \#microcontrollers & 2 & 2 & 4 & 4\\
        \hline
        \#sensors & 6 & 12 & 24 & 48\\
        \hline
        Estimated state-space size& $10^{20}$ & $10^{40}$ & $10^{60}$ & $10^{80}$\\
        \hline
        TTF analysis time (sec) & 20.8 & 33.3 & 49.6 & 89.7 \\
        \hline 
        Conditional TTF analysis time (sec) & 82.9 & 122.6 & 201.3 & 449.1 \\
    \end{tabular}
    \caption{Running time of the benchmark model simulations in seconds}
    \label{tab:run-time-measurements}
\end{table}

The results of the run time measurements are depicted in Table~\ref{tab:run-time-measurements}. Note that all analysis scripts run successfully within 10 minutes.

\section{Conclusion}\label{sec:conclusion}
The top-down analysis of adaptive critical embedded automotive systems is a challenging task due to the high redundancy in the hardware and the applied complex reconfiguration strategies in the software. Traditional modeling and analysis methods do not support the sensor fusion algorithms, the fail-operational adaptation, and the dependent failures. In this work, we introduced a new top-down analysis approach based on the Gamma framework and extended it with stochastic distributions. For the analysis, we created a \textit{PRE}, which provides an easy-to-use interface even for complex analysis techniques, e.g., SVI. Finally, we applied the implemented algorithms on a power-steering model from the automotive industry, and the results show that our algorithm scales well even for large models. Subject to future work, we plan to include the analysis of the sensor fusion algorithms and system effects on the hardware failures, where the modeled controller can optionally modify the failure/environmental distributions.

\paragraph{Acknowledgment}

This work was partially supported by the \'UNKP-19-3 New National Excellence Program of the Ministry for Innovation and Technology and the European Union, co-financed by the European Social Fund (EFOP-3.6.2-16-2017-00013). We thank Dr.~Péter Györke and ThyssenKrupp Presta Kft.\ for the fruitful collaboration and the case study.

\bibliographystyle{eptcs}
\bibliography{bib/bib}   

\begin{thebibliography}{10}
\providecommand{\bibitemdeclare}[2]{}
\providecommand{\surnamestart}{}
\providecommand{\surnameend}{}
\providecommand{\urlprefix}{Available at }
\providecommand{\url}[1]{\texttt{#1}}
\providecommand{\href}[2]{\texttt{#2}}
\providecommand{\urlalt}[2]{\href{#1}{#2}}
\providecommand{\doi}[1]{doi:\urlalt{http://dx.doi.org/#1}{#1}}
\providecommand{\bibinfo}[2]{#2}

\bibitemdeclare{techreport}{abernethy1983weibull}
\bibitem{abernethy1983weibull}
\bibinfo{author}{R~B \surnamestart Abernethy\surnameend}, \bibinfo{author}{J~E
  \surnamestart Breneman\surnameend}, \bibinfo{author}{C~H \surnamestart
  Medlin\surnameend} \& \bibinfo{author}{G~L \surnamestart Reinman\surnameend}
  (\bibinfo{year}{1983}): \emph{\bibinfo{title}{{Weibull} Analysis Handbook}}.
\newblock \bibinfo{type}{AFWAL-TR} \bibinfo{number}{83-2079},
  \bibinfo{institution}{Air Force Wright Aeronautical Laboratories}.
\newblock
  \urlprefix\url{https://apps.dtic.mil/dtic/tr/fulltext/u2/a143100.pdf}.

\bibitemdeclare{article}{andrieu2010particle}
\bibitem{andrieu2010particle}
\bibinfo{author}{Christophe \surnamestart Andrieu\surnameend},
  \bibinfo{author}{Arnaud \surnamestart Doucet\surnameend} \&
  \bibinfo{author}{Roman \surnamestart Holenstein\surnameend}
  (\bibinfo{year}{2010}): \emph{\bibinfo{title}{Particle Markov chain Monte
  Carlo methods}}.
\newblock {\sl \bibinfo{journal}{Journal of the Royal Statistical Society:
  Series B (Statistical Methodology)}}
  \bibinfo{volume}{72}(\bibinfo{number}{3}), pp. \bibinfo{pages}{269--342},
  \doi{10.1111/j.1467-9868.2009.00736.x}.

\bibitemdeclare{techreport}{reliability}
\bibitem{reliability}
\bibinfo{author}{Algirdas \surnamestart Avizienis\surnameend},
  \bibinfo{author}{Jean-Claude \surnamestart Laprie\surnameend},
  \bibinfo{author}{Brian \surnamestart Randell\surnameend} \&
  \bibinfo{author}{Carl \surnamestart Landwehr\surnameend}
  (\bibinfo{year}{2004}): \emph{\bibinfo{title}{Basic Concepts and Taxonomy of
  Dependable and Secure Computing}}.
\newblock \bibinfo{type}{Technical Report} \bibinfo{number}{2004-47},
  \bibinfo{institution}{University of Maryland}.
\newblock
  \urlprefix\url{https://drum.lib.umd.edu/bitstream/handle/1903/6459/TR_2004-47.pdf}.

\bibitemdeclare{article}{bingham2019pyro}
\bibitem{bingham2019pyro}
\bibinfo{author}{Eli \surnamestart Bingham\surnameend},
  \bibinfo{author}{Jonathan~P. \surnamestart Chen\surnameend},
  \bibinfo{author}{Martin \surnamestart Jankowiak\surnameend},
  \bibinfo{author}{Fritz \surnamestart Obermeyer\surnameend},
  \bibinfo{author}{Neeraj \surnamestart Pradhan\surnameend},
  \bibinfo{author}{Theofanis \surnamestart Karaletsos\surnameend},
  \bibinfo{author}{Rohit \surnamestart Singh\surnameend}, \bibinfo{author}{Paul
  \surnamestart Szerlip\surnameend}, \bibinfo{author}{Paul \surnamestart
  Horsfall\surnameend} \& \bibinfo{author}{Noah~D. \surnamestart
  Goodman\surnameend} (\bibinfo{year}{2019}): \emph{\bibinfo{title}{Pyro: Deep
  Universal Probabilistic Programming}}.
\newblock {\sl \bibinfo{journal}{J. Mach. Learn. Res.}}
  \bibinfo{volume}{20}(\bibinfo{number}{1}), p. \bibinfo{pages}{973–978},
  \doi{10.5555/3322706.3322734}.

\bibitemdeclare{inproceedings}{bulychev2012uppaal}
\bibitem{bulychev2012uppaal}
\bibinfo{author}{Peter \surnamestart Bulychev\surnameend},
  \bibinfo{author}{Alexandre \surnamestart David\surnameend},
  \bibinfo{author}{Kim~Gulstrand \surnamestart Larsen\surnameend},
  \bibinfo{author}{Marius \surnamestart Miku{\v{c}}ionis\surnameend},
  \bibinfo{author}{Danny~B{\o}gsted \surnamestart Poulsen\surnameend},
  \bibinfo{author}{Axel \surnamestart Legay\surnameend} \&
  \bibinfo{author}{Zheng \surnamestart Wang\surnameend} (\bibinfo{year}{2012}):
  \emph{\bibinfo{title}{{UPPAAL-SMC}: Statistical model checking for priced
  timed automata}}.
\newblock In: {\sl \bibinfo{booktitle}{QAPL 2012}}, {\sl \bibinfo{series}{Elec.
  Proc. Theor. Comput. Sci.}}~\bibinfo{volume}{85}, pp. \bibinfo{pages}{1--16},
  \doi{10.4204/EPTCS.85.1}.

\bibitemdeclare{article}{burton2003innovationepas}
\bibitem{burton2003innovationepas}
\bibinfo{author}{A.~W. \surnamestart {Burton}\surnameend}
  (\bibinfo{year}{2003}): \emph{\bibinfo{title}{Innovation drivers for electric
  power-assisted steering}}.
\newblock {\sl \bibinfo{journal}{IEEE Control Systems Magazine}}
  \bibinfo{volume}{23}(\bibinfo{number}{6}), pp. \bibinfo{pages}{30--39},
  \doi{10.1109/MCS.2003.1251179}.

\bibitemdeclare{article}{carpenter2017stan}
\bibitem{carpenter2017stan}
\bibinfo{author}{Bob \surnamestart Carpenter\surnameend},
  \bibinfo{author}{Andrew \surnamestart Gelman\surnameend},
  \bibinfo{author}{Matthew~D \surnamestart Hoffman\surnameend},
  \bibinfo{author}{Daniel \surnamestart Lee\surnameend}, \bibinfo{author}{Ben
  \surnamestart Goodrich\surnameend}, \bibinfo{author}{Michael \surnamestart
  Betancourt\surnameend}, \bibinfo{author}{Marcus \surnamestart
  Brubaker\surnameend}, \bibinfo{author}{Jiqiang \surnamestart Guo\surnameend},
  \bibinfo{author}{Peter \surnamestart Li\surnameend} \& \bibinfo{author}{Allen
  \surnamestart Riddell\surnameend} (\bibinfo{year}{2017}):
  \emph{\bibinfo{title}{{Stan}: A probabilistic programming language}}.
\newblock {\sl \bibinfo{journal}{J. Stat. Softw.}}
  \bibinfo{volume}{76}(\bibinfo{number}{1}), \doi{10.18637/jss.v076.i01}.

\bibitemdeclare{inproceedings}{chen2014stochastichmc}
\bibitem{chen2014stochastichmc}
\bibinfo{author}{Tianqi \surnamestart Chen\surnameend}, \bibinfo{author}{Emily
  \surnamestart Fox\surnameend} \& \bibinfo{author}{Carlos \surnamestart
  Guestrin\surnameend} (\bibinfo{year}{2014}): \emph{\bibinfo{title}{Stochastic
  Gradient Hamiltonian Monte Carlo}}.
\newblock In \bibinfo{editor}{Eric~P. \surnamestart Xing\surnameend} \&
  \bibinfo{editor}{Tony \surnamestart Jebara\surnameend}, editors: {\sl
  \bibinfo{booktitle}{Proceedings of the 31st International Conference on
  Machine Learning}}, {\sl \bibinfo{series}{Proceedings of Machine Learning
  Research}}~\bibinfo{volume}{32}, \bibinfo{publisher}{PMLR},
  \bibinfo{address}{Bejing, China}, pp. \bibinfo{pages}{1683--1691}.

\bibitemdeclare{article}{dugan2000developingdfta}
\bibitem{dugan2000developingdfta}
\bibinfo{author}{Joanne~Bechta \surnamestart Dugan\surnameend},
  \bibinfo{author}{Kevin~J \surnamestart Sullivan\surnameend} \&
  \bibinfo{author}{David \surnamestart Coppit\surnameend}
  (\bibinfo{year}{2000}): \emph{\bibinfo{title}{Developing a low-cost
  high-quality software tool for dynamic fault-tree analysis}}.
\newblock {\sl \bibinfo{journal}{IEEE Trans. Reliab.}}
  \bibinfo{volume}{49}(\bibinfo{number}{1}), pp. \bibinfo{pages}{49--59},
  \doi{10.1109/24.855536}.

\bibitemdeclare{inproceedings}{filipovikj2016simulinkautomotive}
\bibitem{filipovikj2016simulinkautomotive}
\bibinfo{author}{Predrag \surnamestart Filipovikj\surnameend},
  \bibinfo{author}{Nesredin \surnamestart Mahmud\surnameend},
  \bibinfo{author}{Raluca \surnamestart Marinescu\surnameend},
  \bibinfo{author}{Cristina \surnamestart Seceleanu\surnameend},
  \bibinfo{author}{Oscar \surnamestart Ljungkrantz\surnameend} \&
  \bibinfo{author}{Henrik \surnamestart Lönn\surnameend}
  (\bibinfo{year}{2016}): \emph{\bibinfo{title}{Simulink to UPPAAL Statistical
  Model Checker: Analyzing Automotive Industrial Systems}}.
\newblock In: {\sl \bibinfo{booktitle}{FM 2016}}, pp.
  \bibinfo{pages}{748--756}, \doi{10.1007/978-3-319-48989-6_46}.

\bibitemdeclare{inproceedings}{graicsmolnarminisy18}
\bibitem{graicsmolnarminisy18}
\bibinfo{author}{Bence \surnamestart Graics\surnameend} \&
  \bibinfo{author}{Vince \surnamestart Moln{\'a}r\surnameend}
  (\bibinfo{year}{2018}): \emph{\bibinfo{title}{Mix-and-Match Composition in
  the {Gamma Framework}}}.
\newblock In: {\sl \bibinfo{booktitle}{25th Minisymposium, Department of
  Measurement and Information Systems}}, \bibinfo{address}{Budapest, Hungary}.

\bibitemdeclare{article}{Harel:1987:SVF:34884.34886}
\bibitem{Harel:1987:SVF:34884.34886}
\bibinfo{author}{David \surnamestart Harel\surnameend} (\bibinfo{year}{1987}):
  \emph{\bibinfo{title}{Statecharts: A Visual Formalism for Complex Systems}}.
\newblock {\sl \bibinfo{journal}{Sci. Comput. Program.}}
  \bibinfo{volume}{8}(\bibinfo{number}{3}), pp. \bibinfo{pages}{231--274},
  \doi{10.1016/0167-6423(87)90035-9}.

\bibitemdeclare{book}{us1986military}
\bibitem{us1986military}
\bibinfo{author}{J.~W. \surnamestart {Harms}\surnameend}
  (\bibinfo{year}{2010}): \emph{\bibinfo{title}{Revision of MIL-HDBK-217,
  Reliability Prediction of Electronic Equipment}}.
\newblock \doi{10.1109/RAMS.2010.5448046}.

\bibitemdeclare{article}{hoffman2013stochasticsvi}
\bibitem{hoffman2013stochasticsvi}
\bibinfo{author}{Matthew~D. \surnamestart Hoffman\surnameend},
  \bibinfo{author}{David~M. \surnamestart Blei\surnameend},
  \bibinfo{author}{Chong \surnamestart Wang\surnameend} \&
  \bibinfo{author}{John \surnamestart Paisley\surnameend}
  (\bibinfo{year}{2013}): \emph{\bibinfo{title}{Stochastic Variational
  Inference}}.
\newblock {\sl \bibinfo{journal}{J. Mach. Learn. Res.}}
  \bibinfo{volume}{14}(\bibinfo{number}{1}), p. \bibinfo{pages}{1303–1347},
  \doi{10.5555/2567709.2502622}.

\bibitemdeclare{article}{hoffman2014nonuts}
\bibitem{hoffman2014nonuts}
\bibinfo{author}{Matthew~D. \surnamestart Homan\surnameend} \&
  \bibinfo{author}{Andrew \surnamestart Gelman\surnameend}
  (\bibinfo{year}{2014}): \emph{\bibinfo{title}{The No-U-Turn Sampler:
  Adaptively Setting Path Lengths in Hamiltonian Monte Carlo}}.
\newblock {\sl \bibinfo{journal}{J. Mach. Learn. Res.}}
  \bibinfo{volume}{15}(\bibinfo{number}{1}), p. \bibinfo{pages}{1593–1623},
  \doi{10.5555/2627435.2638586}.

\bibitemdeclare{techreport}{iso26262}
\bibitem{iso26262}
\bibinfo{author}{\surnamestart {ISO/TC 22/SC 32}\surnameend}
  (\bibinfo{year}{2018}): \emph{\bibinfo{title}{Road vehicles --- Functional
  safety --- Part 9: Automotive Safety Integrity Level ({ASIL})-oriented and
  safety-oriented analyses}}.
\newblock \bibinfo{type}{ISO} \bibinfo{number}{26262-9:2018},
  \bibinfo{institution}{International Organization for Standardization}.
\newblock \urlprefix\url{https://www.iso.org/standard/51365.html}.

\bibitemdeclare{inproceedings}{Katoen16landscape}
\bibitem{Katoen16landscape}
\bibinfo{author}{Joost-Pieter \surnamestart Katoen\surnameend}
  (\bibinfo{year}{2016}): \emph{\bibinfo{title}{The Probabilistic Model
  Checking Landscape}}.
\newblock In: {\sl \bibinfo{booktitle}{Proceedings of the 31st Annual ACM/IEEE
  Symposium on Logic in Computer Science}}, \bibinfo{series}{LICS ’16},
  \bibinfo{publisher}{Association for Computing Machinery},
  \bibinfo{address}{New York, NY, USA}, p. \bibinfo{pages}{31–45},
  \doi{10.1145/2933575.2934574}.

\bibitemdeclare{inproceedings}{kwiatkowska2011prism}
\bibitem{kwiatkowska2011prism}
\bibinfo{author}{Marta \surnamestart Kwiatkowska\surnameend},
  \bibinfo{author}{Gethin \surnamestart Norman\surnameend} \&
  \bibinfo{author}{David \surnamestart Parker\surnameend}
  (\bibinfo{year}{2011}): \emph{\bibinfo{title}{{PRISM 4.0}: Verification of
  Probabilistic Real-time Systems}}.
\newblock In: {\sl \bibinfo{booktitle}{CAV 2011}}, {\sl \bibinfo{series}{LNCS}}
  \bibinfo{volume}{6806}, \bibinfo{publisher}{Springer}, pp.
  \bibinfo{pages}{585--591}, \doi{10.1007/978-3-642-22110-1_47}.

\bibitemdeclare{inproceedings}{molnaretalicse18}
\bibitem{molnaretalicse18}
\bibinfo{author}{Vince \surnamestart Moln{\'a}r\surnameend},
  \bibinfo{author}{Bence \surnamestart Graics\surnameend},
  \bibinfo{author}{Andr{\'a}s \surnamestart V{\"o}r{\"o}s\surnameend},
  \bibinfo{author}{Istv{\'a}n \surnamestart Majzik\surnameend} \&
  \bibinfo{author}{D{\'a}niel \surnamestart Varr{\'o}\surnameend}
  (\bibinfo{year}{2018}): \emph{\bibinfo{title}{The {Gamma} Statechart
  Composition Framework}}.
\newblock In: {\sl \bibinfo{booktitle}{ICSE 2018}}, \bibinfo{publisher}{ACM},
  pp. \bibinfo{pages}{113--116}, \doi{10.1145/3183440.3183489}.

\bibitemdeclare{article}{perov2019multiverse}
\bibitem{perov2019multiverse}
\bibinfo{author}{Yura~N. \surnamestart Perov\surnameend},
  \bibinfo{author}{Logan \surnamestart Graham\surnameend},
  \bibinfo{author}{Kostis \surnamestart Gourgoulias\surnameend},
  \bibinfo{author}{Jonathan~G. \surnamestart Richens\surnameend},
  \bibinfo{author}{Ciar{\'a}n~M. \surnamestart Lee\surnameend},
  \bibinfo{author}{Adam \surnamestart Baker\surnameend} \&
  \bibinfo{author}{Saurabh \surnamestart Johri\surnameend}
  (\bibinfo{year}{2019}): \emph{\bibinfo{title}{MultiVerse: Causal Reasoning
  using Importance Sampling in Probabilistic Programming}}.
\newblock {\sl \bibinfo{journal}{ArXiv}} \bibinfo{volume}{abs/1910.08091}.

\bibitemdeclare{inproceedings}{ruijters2016betterreawayengineering}
\bibitem{ruijters2016betterreawayengineering}
\bibinfo{author}{Enno \surnamestart Ruijters\surnameend} \&
  \bibinfo{author}{Mari{\"e}lle \surnamestart Stoelinga\surnameend}
  (\bibinfo{year}{2016}): \emph{\bibinfo{title}{Better Railway Engineering
  Through Statistical Model Checking}}.
\newblock In: {\sl \bibinfo{booktitle}{Leveraging Applications of Formal
  Methods, Verification and Validation: Foundational Techniques}},
  \bibinfo{publisher}{Springer}, pp. \bibinfo{pages}{151--165},
  \doi{10.1007/978-3-319-47166-2_10}.

\bibitemdeclare{misc}{2016ascl.soft10016Spymc3}
\bibitem{2016ascl.soft10016Spymc3}
\bibinfo{author}{J.~\surnamestart {Salvatier}\surnameend},
  \bibinfo{author}{T.~V. \surnamestart Wiecki{\^a}\surnameend} \&
  \bibinfo{author}{C.~\surnamestart {Fonnesbeck}\surnameend}
  (\bibinfo{year}{2016}): \emph{\bibinfo{title}{{PyMC3}: Python probabilistic
  programming framework}}.
\newblock \bibinfo{howpublished}{Astrophysics Source Code Library}.
\newblock
  \urlprefix\url{https://ui.adsabs.harvard.edu/abs/2016ascl.soft10016S}.

\bibitemdeclare{inproceedings}{shan2014formalaerospace}
\bibitem{shan2014formalaerospace}
\bibinfo{author}{Lijun \surnamestart Shan\surnameend}, \bibinfo{author}{Yuying
  \surnamestart Wang\surnameend}, \bibinfo{author}{Ning \surnamestart
  Fu\surnameend}, \bibinfo{author}{Xingshe \surnamestart Zhou\surnameend},
  \bibinfo{author}{Lei \surnamestart Zhao\surnameend}, \bibinfo{author}{Lijng
  \surnamestart Wan\surnameend}, \bibinfo{author}{Lei \surnamestart
  Qiao\surnameend} \& \bibinfo{author}{Jianxin \surnamestart Chen\surnameend}
  (\bibinfo{year}{2014}): \emph{\bibinfo{title}{Formal Verification of Lunar
  Rover Control Software Using UPPAAL}}.
\newblock In \bibinfo{editor}{Cliff \surnamestart Jones\surnameend},
  \bibinfo{editor}{Pekka \surnamestart Pihlajasaari\surnameend} \&
  \bibinfo{editor}{Jun \surnamestart Sun\surnameend}, editors: {\sl
  \bibinfo{booktitle}{FM 2014: Formal Methods}}, \bibinfo{publisher}{Springer},
  pp. \bibinfo{pages}{718--732}, \doi{10.1007/978-3-319-06410-9_48}.

\bibitemdeclare{article}{taheriyoun2015reliabilityftsimulation}
\bibitem{taheriyoun2015reliabilityftsimulation}
\bibinfo{author}{Masoud \surnamestart Taheriyoun\surnameend} \&
  \bibinfo{author}{Saber \surnamestart Moradinejad\surnameend}
  (\bibinfo{year}{2015}): \emph{\bibinfo{title}{Reliability analysis of a
  wastewater treatment plant using fault tree analysis and Monte Carlo
  simulation}}.
\newblock {\sl \bibinfo{journal}{Environmental Monitoring and Assessment}}
  \bibinfo{volume}{187}(\bibinfo{number}{1}), p. \bibinfo{pages}{4186},
  \doi{10.1007/s10661-014-4186-7}.

\bibitemdeclare{article}{tran2016edward}
\bibitem{tran2016edward}
\bibinfo{author}{Dustin \surnamestart Tran\surnameend}, \bibinfo{author}{Alp
  \surnamestart Kucukelbir\surnameend}, \bibinfo{author}{Adji~B \surnamestart
  Dieng\surnameend}, \bibinfo{author}{Maja \surnamestart Rudolph\surnameend},
  \bibinfo{author}{Dawen \surnamestart Liang\surnameend} \&
  \bibinfo{author}{David~M \surnamestart Blei\surnameend}
  (\bibinfo{year}{2016}): \emph{\bibinfo{title}{{Edward}: A library for
  probabilistic modeling, inference, and criticism}}.
\newblock {\sl \bibinfo{journal}{arXiv preprint arXiv:1610.09787}}.

\bibitemdeclare{article}{zhang2012reliabilityftsimulation}
\bibitem{zhang2012reliabilityftsimulation}
\bibinfo{author}{P.~\surnamestart {Zhang}\surnameend} \& \bibinfo{author}{K.~W.
  \surnamestart {Chan}\surnameend} (\bibinfo{year}{2012}):
  \emph{\bibinfo{title}{Reliability Evaluation of Phasor Measurement Unit Using
  Monte Carlo Dynamic Fault Tree Method}}.
\newblock {\sl \bibinfo{journal}{IEEE Transactions on Smart Grid}}
  \bibinfo{volume}{3}(\bibinfo{number}{3}), pp. \bibinfo{pages}{1235--1243},
  \doi{10.1109/TSG.2011.2180937}.

\bibitemdeclare{article}{vcepin2002dynamicfaulttreedft}
\bibitem{vcepin2002dynamicfaulttreedft}
\bibinfo{author}{Marko \surnamestart Čepin\surnameend} \&
  \bibinfo{author}{Borut \surnamestart Mavko\surnameend}
  (\bibinfo{year}{2002}): \emph{\bibinfo{title}{A dynamic fault tree}}.
\newblock {\sl \bibinfo{journal}{Reliability Engineering \& System Safety}}
  \bibinfo{volume}{75}(\bibinfo{number}{1}), pp. \bibinfo{pages}{83 -- 91},
  \doi{10.1016/S0951-8320(01)00121-1}.

\end{thebibliography}

\newpage

\setGammaSyntax

\setcounter{section}{0}
\renewcommand{\thesection}{\Alph{section}}

\section{Gamma composition semantics}\label{app:gamma}

Gamma is a modeling framework for the semantically precise composition of statechart components. Statecharts (considered as atomic components) can be composed in the Gamma Composition Language (GCL), which supports the definition of \emph{synchronous} and \emph{asynchronous} composite components, two fundamentally different system types determining how their constituent components receive events and how they are executed. In the subsequent sections, we informally introduce the communication elements in GCL as well as the cascade composition mode of synchronous systems, as we used this composition mode for defining the EPAS configuration. Additional information on the synchronous and cascade composition modes, as well as asynchronous systems, can be found in~\cite{graicsmolnarminisy18}.

\subsection{Communication Elements}
\label{sec:communication-elements}

In GCL, components (both atomic and composite) communicate through \emph{ports}. Each port defines a point of service through which certain \emph{event notifications} can be sent or received. An event notification (or event for short) is a piece of information passed between components, which can also have \emph{parameters} to forward data. An event is called \emph{message} in the case of asynchronous components and \emph{signal} in the case of synchronous components. Events are declared on \emph{interfaces}, which may be realized by ports. An event may be declared as \emph{input} or \emph{output}. The declared directions are reversed, however, if the port does not \emph{provide}, but \emph{require} the interface, which are the two possible modes in which a port can realize an interface.

\subsection{Synchronous components}
Synchronous components represent models that communicate
in a synchronous manner using \emph{signals}. They are executed in a lockstep fashion, triggered by an enclosing component (synchronous or asynchronous) or an external actor from the environment. When executed, synchronous components process incoming signals and produce output signals in accordance with their internal states. Input signals are not queued but sampled: upon execution, the component can access the most recent signal for each event on every port since the last execution (if there is any). Similarly, output signals are reset at the beginning of every execution and each output event on every port can get a new signal assigned to it.

Synchronous components in Gamma are \emph{statechart definitions}, which are considered atomic components as well as \emph{synchronous} and \emph{cascade composite components}, which can be freely mixed in hierarchically composed synchronous systems.

\subsection{Cascade composite components} Conceptually, components in a cascade composite model represent a set of ``filters'' through which inputs are transformed into outputs. Therefore, constituent components immediately see the output signals of other components in the same composite component during execution. By default, constituent components are executed once in the \emph{order of their instantiation}. Alternatively, an execution list can be defined that determines the execution order of instantiated components. The execution list can contain a particular constituent component many times, supporting repeated execution. The typical arrangement of a cascade composite component definition is illustrated in the following snippet.

\begin{minipage}{0.91\linewidth}
\begin{lstlisting}[caption={The textual representation of a typical Gamma cascade composite component}]
cascade Epas [
    // System port declarations
    port S1AFault: requires SensorFault
    // ...
] {
    // Component instances
    component S1A: SensorStatechart
    component DiagA: DiagnosticStatechart
    // ...
    // Binding composite model ports to internal ports
    bind S1AFault->S1A.HWFault
    // ...
    // Channel definitions connecting internal ports:
    // A provided and required realization of the same interface
    channel [S1A.SensorFault] -o)- [DiagA.S1HW]
    // ...
}
\end{lstlisting}
\end{minipage}

\section{Models}

\subsection{System layer}

\newpage

\begin{lstlisting}[caption={Gamma textual model of the EPAS configuration}]
package epas
import "interfaces/Interfaces.gcd"

cascade Epas [
    port State: provides Eval
	port S1AFault: requires SensorFault
	port S2AFault: requires SensorFault
	port S3AFault: requires SensorFault
	port S1BFault: requires SensorFault
	port S2BFault: requires SensorFault
	port S3BFault: requires SensorFault
	port UCAFault: requires UCFault
	port UCBFault: requires UCFault
	
] {
	component S1A: SensorStatechart
	component S2A: SensorStatechart
	component S3A: SensorStatechart
	
	component S1B: SensorStatechart
	component S2B: SensorStatechart
	component S3B: SensorStatechart
	
	component DiagA: DiagnosticStatechart
	component DiagB: DiagnosticStatechart
	
	component UCA: UCStatechart
	component UCB: UCStatechart
	
	component ACTRL: MainctrlStatechart
	component BCTRL: MainctrlStatechart
	
	component Ev: EvaluationStatechart
	
	bind S1AFault->S1A.HWFault
	bind S2AFault->S2A.HWFault
	bind S3AFault->S3A.HWFault
	
	bind S1BFault->S1B.HWFault
	bind S2BFault->S2B.HWFault
	bind S3BFault->S3B.HWFault
	
	bind UCAFault->UCA.HWFault
	bind UCBFault->UCB.HWFault
	
	bind State->Ev.Eval
	
	channel [S1A.SensorFault] -o)- [DiagA.S1HW]
	channel [S2A.SensorFault] -o)- [DiagA.S2HW]
	channel [S3A.SensorFault] -o)- [DiagA.S3HW]
	
	channel [S1B.SensorFault] -o)- [DiagB.S1HW]
	channel [S2B.SensorFault] -o)- [DiagB.S2HW]
	channel [S3B.SensorFault] -o)- [DiagB.S3HW]
	
	channel [DiagA.DiagnosticOutput] -o)- [ACTRL.DiagnosticOutput]
	channel [DiagA.DiagnosticStatus] -o)- [ACTRL.DiagnosticStatus]
	channel [DiagB.DiagnosticOutput] -o)- [BCTRL.DiagnosticOutput]
	channel [DiagB.DiagnosticStatus] -o)- [BCTRL.DiagnosticStatus]
	
	channel [UCA.Fault] -o)- [ACTRL.UCHW]
	channel [UCB.Fault] -o)- [BCTRL.UCHW]
	
	channel [ACTRL.Monitor] -o)- [Ev.AMonitor]
	channel [BCTRL.Monitor] -o)- [Ev.BMonitor]
}
\end{lstlisting}

\begin{figure}[!h]
	\noindent\begin{minipage}{.47\textwidth}
	\begin{lstlisting}[frame=tlrb]
interface UCFault {
	out event shutdown
}

interface SensorFault {
	out event det
	out event latent
}

interface DiagnosticStatus {
	out event Error
	out event Warning
}
\end{lstlisting}
	\end{minipage}\hfill
	\begin{minipage}{.47\textwidth}
		\begin{lstlisting}[frame=tlrb]
interface Eval {
	out event SS
	out event SLoA
}

interface DiagnosticOutput {
	out event WrongOutput
}

interface Monitor {
	out event warning
	out event loa
	out event selfsteering
}
\end{lstlisting}
	\end{minipage}
	\caption{Gamma textual models of the EPAS interfaces}
	\label{snippet:modes3-interfaces}
\end{figure}

\newpage

\subsection{Safety layer}

\newpage

\subsection{Hardware layer}

\begin{table}[ht!]
    \centering\begin{tabular}{c|c|c|c|c}
        Failure distribution & Distribution parameters & HW statechart & From state & To state \\
        \hline
        \hline
        Weibull & \(\begin{array}{@{}c@{}}\text{concentration=1.5,}\\\text{scale=0.1$\cdot 10^{-9}\frac{1}{h}$}\end{array}\)  & uC & On & Off \\
        \hline
        Exponential & rate=10.0 $\cdot 10^{-9}\frac{1}{h}$ & Sensor & Ok & Off \\
        \hline
        Exponential & rate=1.0 $\cdot 10^{-9}\frac{1}{h}$ & Sensor & Ok & LatentFailure 
    \end{tabular}
    \caption{Connection between the distributions and the state transitions in the case-study}\label{tab:faultdistributions}
\end{table}

\section{Probabilistic Runtime Environment}
\label{sec:algorithm}

\begin{algorithm}
\caption{Pseudo code of the translated probabilistic program}
\label{alg:generatefaultseriespseudostatechart}
\begin{algorithmic}
\STATE faults $\gets$ $\emptyset$ 
\FORALL{component $\in$ system}
    \FORALL{failure\_mode $\in$ component}
        \STATE fault\_time $\gets$ sample(failure\_mode.distribution)
        
        \STATE faults $\gets \{\langle\text{fault_time};\text{component};\text{failure\_mode}\rangle\} \cup \text{faults}$
    \ENDFOR
\ENDFOR
\STATE faults $\gets$ orderByTime(faults)
\STATE state $\gets$ "Normal"
\WHILE{state == "Normal"}
    \STATE fault $\gets$ faults.next()
    \STATE state $\gets$ GammaModel.getResultOf(fault.component, fault.failure_mode)
\ENDWHILE
\RETURN \{fault.time; state\}
\end{algorithmic}
\end{algorithm}

\begin{lstlisting}[caption={Probabilistic program for Weibull distribution fitting with SVI},language=Python,label=lst:weibull]
# Instantiate the generated Probabilistic Runtime Environment
simulate = create_pru()
# The guide function encodes the fitted distribution
def guide():
    scale = pyro.param("scale", torch.tensor(1.0), constraint=constraints.positive)
    shape = pyro.param("shape", torch.tensor(1.0), constraint=constraints.positive)
    pyro.sample("failure_time", dist.Weibull(scale, shape))
# Set up SVI with the Adam optimizer
optimizer = pyro.optim.Adam({"lr": 0.05, "betas": (0.9, 0.999)})
svi = pyro.infer.SVI(simulate, guide, optimizer, loss=pyro.infer.Trace_ELBO())
n_steps = 500
for step in range(n_steps):
    svi.step()
# Extract numerical solution
scale = pyro.param("scale").item()
shape = pyro.param("shape").item()
\end{lstlisting}

\begin{lstlisting}[caption={Probabilistic program for conditional analysis},language=Python,label=lst:conditional]
# Instantiate the generated Probabilistic Runtime Environment
simulate = create_pru()
# Failure mode 0 (passed to torch.tensor) corresponds to self-steering
conditional = pyro.condition(simulate, {"failure_mode": torch.tensor(0.0)})
sampler = pyro.infer.Importance(conditional, num_samples=10000)
# Generate a histogram of the conditional distribution
marginal_dist = pyro.inter.EmpiricalMarginal(sampler.run(), sites="failure_time")
\end{lstlisting}

\end{document}